\documentclass[12pt]{article}
\usepackage[usenames,dvips]{pstcol}
\usepackage{color}
\usepackage{amssymb}
\usepackage{epsfig}
\usepackage{footnote}
\usepackage{longtable}
\usepackage{subfigure}
\usepackage{verbatim}
\usepackage{graphicx}
\usepackage{url}
\usepackage{epstopdf}
\usepackage{setspace}
\usepackage{lineno}

\def\HGMTTM{HGMT\textsuperscript{TM}}

\def\LMCPTM{LMCP\textsuperscript{TM}}

\def\18F{^{18}F}

\def\detector_claim_1{Claim 1 of Section~\ref{claims_detector}}
\newrgbcolor{maroon}{.45 0.1 0.1}
\begin{document}

\pagestyle{plain}
%\renewcommand{\headrulewidth}{0pt}
%
% Version
%
\begin{flushright}
Version v10a (accepted manuscript)\\
\today
\end{flushright}
%==================================================================================
%==================================================================================

% \doublespacing

% Title
%
\begin{center}
{\Large\bf Low-Dose TOF-PET Based on Surface Electron Production in Dielectric Laminar MCPs }\\

% (HGMT)}
\vspace*{0.25in}

Kepler Domurat-Sousa, Cameron Poe, Henry J. Frisch\\
{\it Enrico Fermi Institute, University of Chicago}\\
\vspace*{0.05in}

Bernhard W. Adams\\
{\it Quantum Optics Applied Research}\\
\vspace*{0.05in}

Camden Ertley\\
{\it SouthWest Research Institute}\\
\vspace*{0.05in}

 Neal Sullivan\\
{\it Angstrom Research, Inc}\\
\vspace*{0.05in}

\vskip 0.1in
 {\it  Published in Nuclear Instruments and Methods, Section A}
\end{center}

\vskip-0.5in
% make \paragraph a subsubsubsection and \subparagraph a subsubsubsubsec
%
\setcounter{secnumdepth}{5} % 3 is subsubsection 4 is paragraph
\setcounter{tocdepth}{5}
%===============================================================
%\newpage
% \tableofcontents
% \newpage
%
%\listoffigures
%
%===============================================================
%\vskip 0.15in
%\newpage

\begin{abstract}

We present simulations of whole-body low-dose time-of-flight positron emission tomography (TOF-PET) based on the
direct surface production~\cite{MGM_NIM_paper} by 511 keV gamma rays of
energetic electrons via the Photo-electric and Compton Effects,
eliminating the scintillator and photodetector sub-systems in PET
scanners.  In Ref.~\cite{MGM_NIM_paper} we described Microchannel
Plates (MCP) constructed from thin dielectric laminae containing heavy
nuclei such as lead or tungsten (\LMCPTM). The laminae surfaces are
micro-patterned to form channels, which can then be functionalized to
support secondary electron emission in the manner of conventional MCPs.
After assembly of the laminae, the channels form the pores of the
conversion LMCP. This conversion stage is then followed by a high-gain
MCP-based amplification stage, which also can be constructed using the
laminar technique, but with pores typical of currently-available
large-area MCPs .

 We have simulated direct conversion  using modifications to the TOPAS
Geant4-based tool kit. A 20 $\times$ 20 $\times$ 2.54 cm$^3$ LMCP,
composed of 150-micron thick lead-glass laminae,  is predicted to have
a $\ge 30$\% conversion efficiency to a primary electron that
penetrates an interior wall of a pore. The subsequent secondary
electron shower is largely confined to one pore and can provide high
space and time resolutions.

In whole-body PET scanners the technique eliminates the scintillator
and pho\-to\-de\-tec\-tor subsystems. The consequent absence of a photocathode
allows assembly of large arrays at atmospheric pressure and less
stringent vacuum requirements, including use of pumped and cycled
systems.

TOPAS simulations of the Derenzo and XCAT-brain phantoms are presented
with dose reductions of factors of 100 and 1000 from a literature
benchmark. New ap\-pli\-ca\-tions of PET at a significantly lower radiation
dose include routine screening for early detection of pathologies, the
use in diagnostics in previously unserved patient populations such as
children, and a larger installed facility base in rural and
under-served populations, where simpler gamma detectors and lower
radiation doses may enable small low-cost portable PET scanners.

\end{abstract}
%==================================================================================
%==================================================================================
\newpage

\section{Introduction}

Positron emission tomography (PET) uses radioactive positron-emitting
tracers to locate areas of high biological activity such as tumors and
hair-line fractures of bones. It com\-ple\-ments other modalities that
identify morphologies, and is often used in conjunction with CT or MRI.
In addition PET is used on small animals for development of
pharmaceuticals and treatments.

In the last decade detectors and techniques for time-of-flight
positron emission to\-mog\-ra\-phy (TOF-PET) have substantially grown in
sophistication~\cite{Vandenberghe_Moskal_Karp_review_2020,Vaquero_Kinehan_review_2015,Phelps_Cherry_Dahlbom_book_2006}.
Among other innovations, high-precision whole-body scanners have been
built and
characterized~\cite{Vandenberghe_Moskal_Karp_2020_Whole_Body_PET_2020,Cherry_Explorer_scattering_2019};
TOF-PET with sub-nanosecond coincidence has
recently been developed~\cite{Lee_Levin_100ps_2021}; an international competition to develop sub-10
ps TOF resolution is now in place~\cite{LeCoq_2019_case,LeCoq_2020_10ps_challenge}; and timing with resolutions of 10 ps or below using Cherenkov light in
pre-radiators~\cite{Credo,Ohshima,Anatoly_TestBeam_2010} is being
developed by Cherry et al. for higher spatial resolutions and lower
doses~\cite{Cherry_Hamamatsu_2021}.

Recently, alternative methods have been
proposed~\cite{Eric_CPAD_talk,PET_2021_NIM_paper,PET_2023_TMI_paper,Allison_JLAB_talk}
with the goal of achieving resolutions set by the underlying physics
processes rather than by the detector
segmentation~\cite{Moses_fundamental_limits}.  The technique, which
like the conventional technique uses the conversion of the gamma rays
in a scintillator followed by photo-detection, is to exploit Compton
Scattering of the gamma rays in low atomic number (Z) scintillating media.
Successive Compton scatters are constrained by the two-body Compton
kinematics, allowing precisely locating the first scatter in a large
fraction of events~\cite{PET_2023_TMI_paper,Allison_JLAB_talk}.

With similar motivation, here we have adapted the TOPAS Geant4-based
frame\-work~\cite{TOPAS_Methods_paper,TOPAS,TOPAS_user_support} to study
direct surface conversion of gamma rays to electrons via the Compton
and Photoelectric effects in MCPs constructed from thin micro-patterned
laminae (\LMCPTM) containing heavy nuclei such as lead or tungsten
\cite{MGM_NIM_paper}. The direct conversion technique eliminates the
scintillator and photodetector subsystems in TOF-PET scanners,
converting the gamma ray to an electron shower inside an MCP-based
planar vacuum tube, the High-resolution Gamma Multiplier Tube
(\HGMTTM). In addition to the savings in cost, complexity, and bulk
from not using heavy crystals and photodetectors, the absence of a
photocathode allows assembly of large arrays at atmospheric pressure
and much relaxed vacuum requirements, including use of pumped and
cycled systems.

The organization of the paper is as follows. Section~\ref{HGMT}
introduces the HGMT and its components. Section~\ref{simulation_results} presents gamma ray conversion
efficiencies and resolutions from TOPAS sim\-u\-la\-tions of the LMCP. 
Section~\ref{whole_body_scanner_simulation_results} presents images
from simulations of the Derenzo \cite{Derenzo_phantom} and
XCAT-brain~\cite{XCAT2010paper} phantoms in a whole-body HGMT-based TOF-PET
detector at reduced dose. Portable and animal TOF-PET scanners are discussed in Section \ref{sec:portable-animal-pet}.
Section~\ref{summary_and_conclusions}
summarizes the results of this first software study and recommends
starting to build and test LMCP/HGMT prototypes. Appendix A discusses
future studies of time resolution that are beyond the current scope.
%==================================================================================
%\newpage
%==================================================================================

\section{The High\--resolution Gamma Mul\-ti\-pli\-er Tube \\ (HGMT\textsuperscript{TM})}
\label{HGMT}

The HGMT  is a large-area ($\ge 100$~ cm$^2$), high-gain ($
10^6-5\times 10^7$), low-noise MCP-based electron multiplier vacuum
tube, designed to provide correlated high-resolution space/time
measurements of gamma rays via the technique of surface direct
conversion~\cite{MGM_NIM_paper}. The HGMT consists of two stages: a
conversion stage in which gamma rays interact near a substrate surface
with high-Z nuclei to produce `primary' electrons,  followed by a
high-gain electron amplification stage with high spatial and time
resolution. The implementation of the con\-ver\-sion stage is described
here using microchannel plates (MCP) constructed by stacking thin
patterned laminae (LMCP)~\cite{MGM_NIM_paper}. The amplification stage
can be constructed with LMCPs or conventional MCPs, or can be
integrated towards the downstream end of the pores of the conversion
LMCP, for example.

%
% Figure 1 Surface Direct Production with straight pores
%
\begin{figure}[!th]
\centering
\includegraphics[angle=0,width=0.80\textwidth]{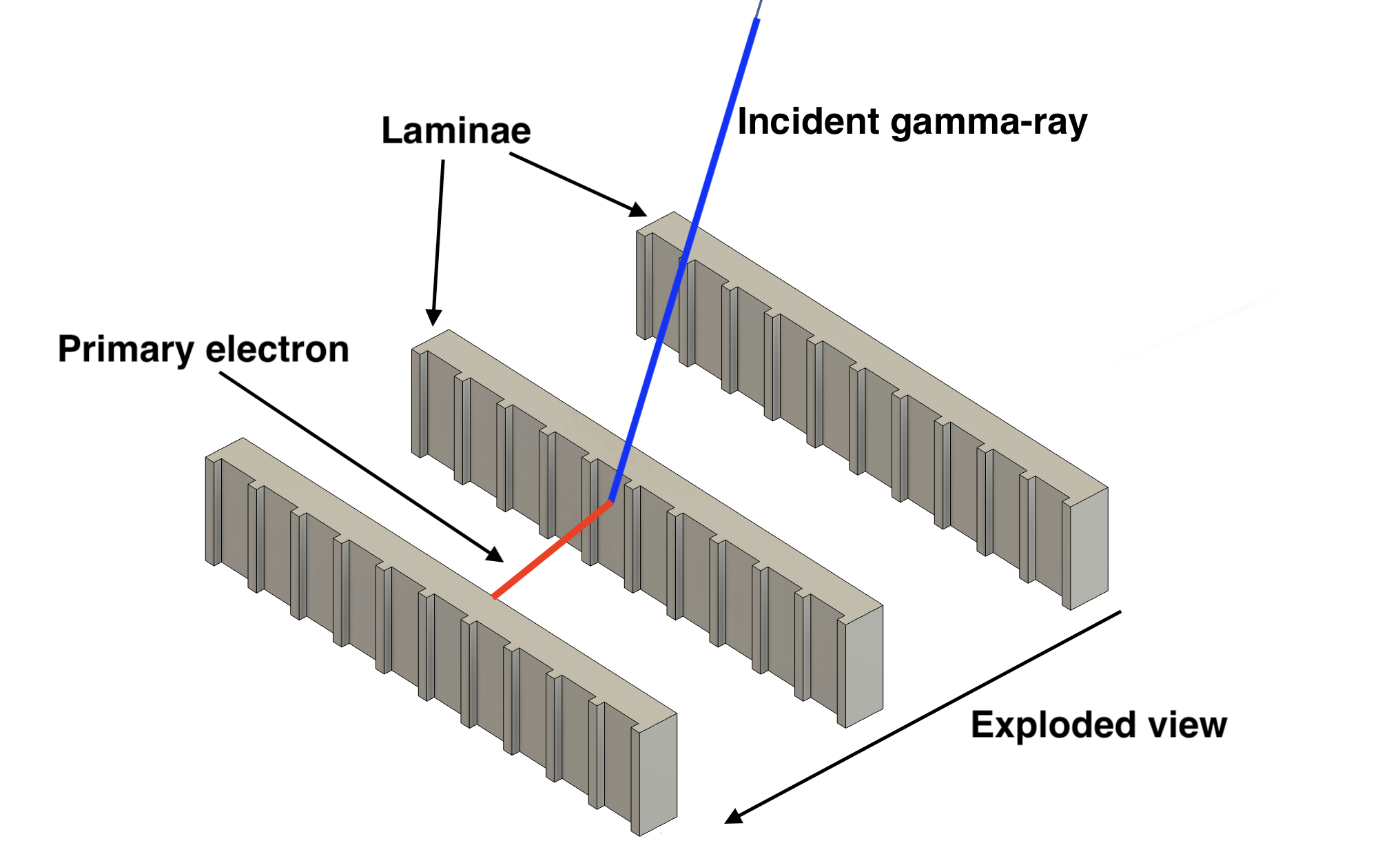}
\caption{The principle of Surface Direct
Conversion~\cite{MGM_NIM_paper}, in which the gamma ray directly
converts to an electron near a surface of an MCP substrate formed from
thin laminae (LMCP) containing high-Z nuclei. Three lamina of the LMCP
are shown, with purely-illustrative straight rectangular pores (not to
scale). This `primary' electron crosses the functionalized surfaces of
a pore, producing secondary electrons in the pore.}
\label{surface_direct_conversion_straight_channel}
\end{figure}

Figure~\ref{surface_direct_conversion_straight_channel} illustrates the
technique.  For a 511 keV gamma ray, the primary electron has a short
range, and must be produced close to a surface to be detected; the
range, typically on the order of a hundred microns in Pb-glass, determines the
thickness of the pore walls. The figure illustrates the process for an
implementation of the HGMT with LMCPs. The dimensions are not to
scale; also the pore geometry and the surface resistive, insulating,
and metal coatings, shown here as uniformly resistive, can be tailored
to provide local well-defined strike surfaces and high-gain
acceleration or low-field drift sections, for
example~\cite{MGM_NIM_paper}.

\subsection{The Laminar Micro-Channel Plate (LMCP)}

%
% Figure 2 LMCP with ridges
%
\begin{figure}[!th]
\centering
\includegraphics[angle=0,width=1.00\textwidth]{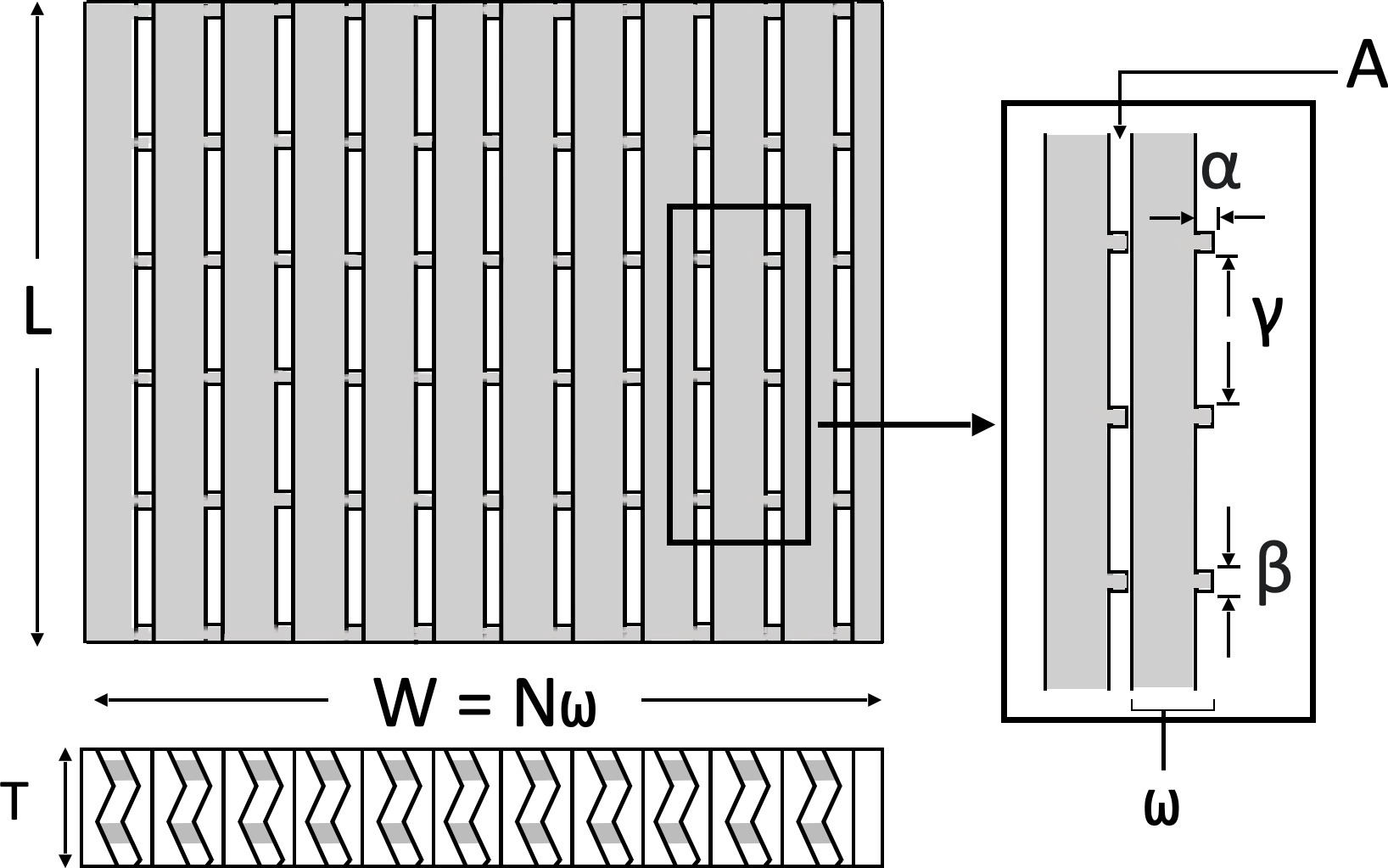}
\caption{A simplified example of an LMCP body intended for gamma ray
detection. The thickness of the body is equal to the width of a lamina,
$T$. Laminae are stacked on edge such that the ridges form pores
between the lamina of dimension $\alpha$ perpendicular to the laminate
and $\gamma$ along the laminate surface (see the right-hand inset). The
pattern of the pore is purely for illustration.}
\label{fig:slab_w_ridges}
\end{figure}

The LMCP, which is a Micro-Channel Plate (MCP) constructed from thin
laminae, is described in detail in Ref.~\cite{MGM_NIM_paper}.
Figure~\ref{fig:slab_w_ridges} shows an example of an LMCP body made
from  a heavy-metal dielectric such as lead-glass~\cite{MGM_NIM_paper}.
The bulk laminae, which represent the largest fraction of the area of
the LMCP, serve to convert the incoming gamma ray to an electron. The
patterning and coating of the open-face laminae surfaces before
assembly allow optimizations in the pore transverse and longitudinal
shapes, variable longitudinal resistance, voltages between strikes, and
pore entrance and exit geometries.\footnote{As described in Ref.~\cite{MGM_NIM_paper}, a funnel entrance can provide an OAR close to 100\% for an
amplification LMCP with 5-10 $\mu$m patterned pores.}

\subsection{The HGMT Components}

%
% Figure 3 HGMT Stackup w Conversion and Amplification MCP
%
\begin{figure}[!th]
\centering
\includegraphics[angle=0,width=1.0\textwidth]{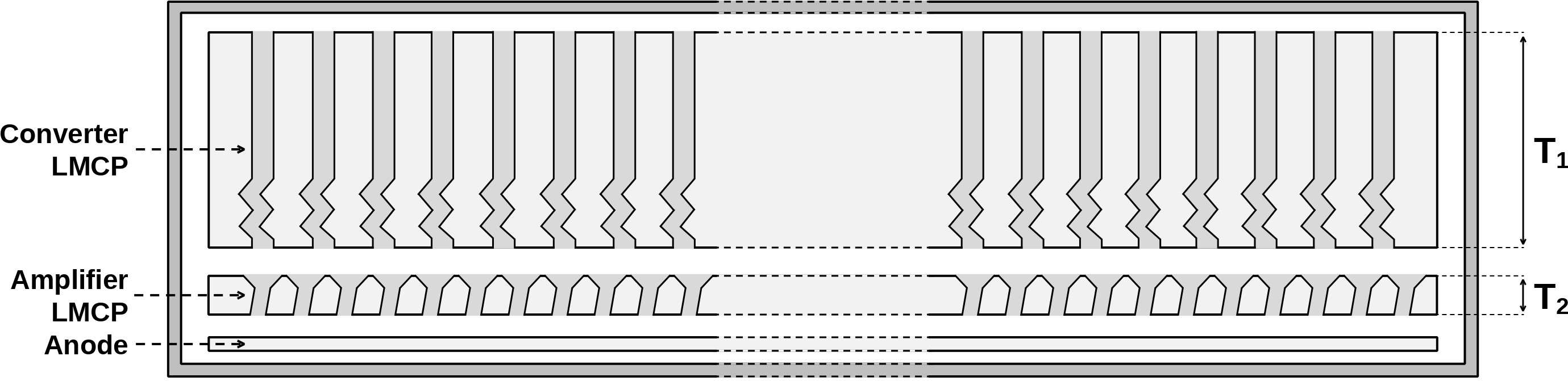}
\caption{An example HGMT detector assembly with a conversion LMCP, an
amplifier LMCP~\cite{MGM_NIM_paper}, and a multi-channel anode.  Gamma
rays are incident from above. Note the aspect ratio of the image is
distorted by the elision of the center of the tube.}
\label{fig:2_slab_millichannel_detector}
\end{figure}

Figure~\ref{fig:2_slab_millichannel_detector} shows a sketch of a
detector assembly of two LMCPs formed with laminae that form structured
pores~\cite{MGM_NIM_paper}. The open-area ratio (OAR) of the converter
LMCP is by design small to maximize the area presented to gamma rays
for direct conversion to electrons. The path length of the gamma rays
in the substrate material depends on the OAR, the thickness and
structure of the laminae, and the incident angles of the gamma ray from
the normal. The amplification stage, either an LMCP or a conventional
MCP, can be made from light glass, and typically will have pores with
diameters of 5-40 $\mu$m and a length-to-diameter (L/D) ratio of 60-80.
The LMCP assembly is followed by an application-specific high-bandwidth
anode with sub-mm resolution for
readout~\cite{Tang_Naxos,RFstrip_anode_paper,patterned_anode_paper}.
Multiple LMCP/anode assemblies can be stacked vertically or
horizontally in a common vacuum vessel.

\subsection{Anode Configurations}
\label{anode_stage}

The multi-channel anode records position and time-of-arrival of the
electron shower after amplification. There are a number of options for
the shape, size, and electrical characteristics of each anode element
(e.g. strip, pad) of a patterned anode depending on the application.
For low-rate environments such as low-dose TOF-PET\footnote{We note
that pile-up of gamma rays from multiple e$^+$e$^-$ annihilations has a
quadratic dependence on rate; a dose reduction of a factor of $10^2$
reduces pile-up by a factor of $10^4$.} there is extensive experience
with anodes with 50-ohm striplines that give sub-mm resolution in both
transverse
directions~\cite{Tang_Naxos,RFstrip_anode_paper,Oberla_thesis,OTPC_paper,Evan_thesis}.
For large-area detectors, such as at a particle collider, arrays of
2-dimensional pads patterned to enhance charge sharing may provide
better resolution than rectangular pads, allowing a large decrease in
electronic channel count while maintaining the same spatial
resolution~\cite{patterned_anode_paper}. For decoupling the anode
design from the tube design, as would be economical for mass production
of single-design HGMT modules for different uses, an option is
capacitive-coupling through an adequately resistive bottom plate for
both stripline and pad anodes~\cite{InsideOut_paper}. Additionally,
anodes can be made from solid-state devices.

\subsection{Vacuum and Hermetic Packaging}
\begin{figure}[!th]
\centering
\includegraphics[angle=0,width=0.52\textwidth]{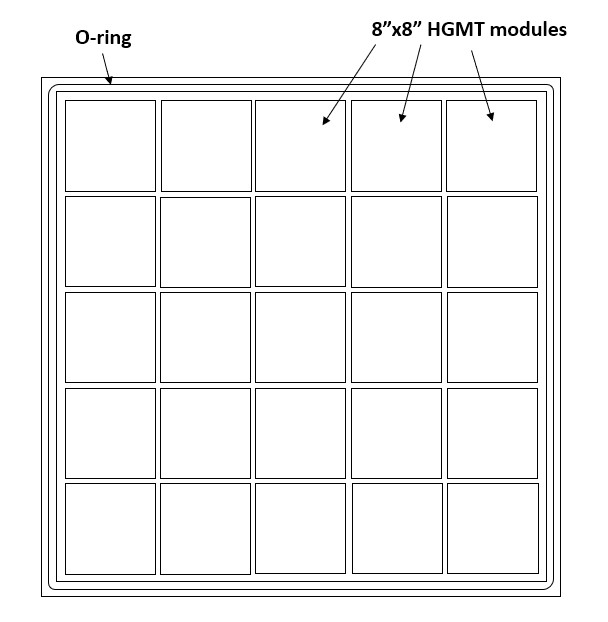}
\hfil
\raisebox{0.15in}{\includegraphics[angle=0,width=0.45\textwidth]{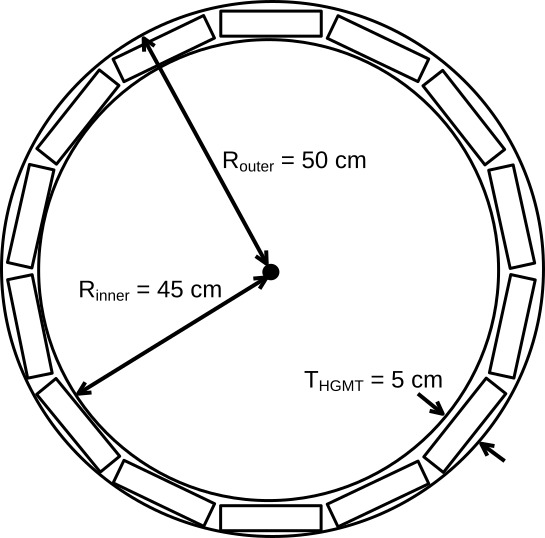}}
%\raisebox{0.6in}{\includegraphics[angle=0,width=0.63\textwidth]{5x5_array_elevation_v1b.jpg}}
\caption{Left: A 5-by-5 array of HGMTs such as would be used in a
pre-converter in a particle physics experiment in a common planar
vacuum package. Right: An axial view of a whole-body PET-TOF scanner
with planar HGMT modules in a common cylindrical vacuum vessel.}
\label{fig:5x5_array}
\end{figure}

The left-hand panel of Figure~\ref{fig:5x5_array} represents a  5-by-5
array of HGMTs in a common planar vacuum package such as might be used in a rare kaon decay experiment or at high rapidity in a collider experiment. For 20
cm (8")-square modules the array covers approximately 1 m$^2$.
Strip-line anodes with sub-mm resolution may be shared across the full
extent~\cite{timing_paper}, lowering the electronics channel count if
rates allow. The right-hand panel is an axial sketch of a PET-scanner
array comprised of planar\footnote{In practice these would most likely
be sections of a cylinder~\cite{MGM_NIM_paper}.} LMCP modules that
replace the scintillator and photo-detector systems. Multiple modules,
including up to the complete array, can share a vacuum vessel.
%
% Figure 4
%

The LMCP construction allows wide flexibility in shape and materials
for the vacuum package. For gamma rays the tube body can be metal as
well as the conventional glass or ceramic. The package may be
non-rectangular or non-planar to fit non-standard shaped
LMCPs~\cite{MGM_NIM_paper}. Appropriately spaced tabs on the perimeter
of the laminae can provide support against atmospheric pressure from
top to bottom of the tube. As the HGMT has no photocathode, vacuum
sealing can be done at atmospheric pressure, with a higher target
pressure. Sealing with O-rings, active pumping, and cycling to
atmospheric pressure for maintenance or transport become options.

Large systems of HGMTs may be installed in a single vessel such as a
cylindrical vessel with an open bore for a PET subsystem, or a large
planar vacuum vessel for a photon/electron pre-sampler in a particle
physics experiment. For some applications, such as in a large particle
collider experiment or inside the magnet bore in a multi-modality PET
detector, the HGMT thin aspect ratio saves expensive real-estate over
crystal-based gamma ray detection systems. Common packaging also enables a higher packing fraction and
economies of shared subsystems. Large systems can be continuously
pumped rather than sealed, and can be brought up to atmospheric
pressure for maintenance or modification.

\section{Simulation Results: Con\-ver\-sion Ef\-fi\-cien\-cies and \\ Space and Time Resolutions}
\label{simulation_results}

The analyses of the efficiencies and resolutions of the HGMT were
carried out using sim\-u\-la\-tions performed with TOPAS \cite{TOPAS}, a
user-friendly interface for the Geant4 simulation package of the
interaction of particles with matter \cite{Geant4_2003,Geant4_2010}.
The simulations of gamma rays, positrons, and electrons used the Geant4
physics lists ``G4EMStandardPhysics\_option4'' and \\
``G4EmPenelopePhysics'' \cite{geant4_physics_lists}. The substrate material for the HGMT was set as the internal Geant4 material ``G4\_GLASS\_LEAD''.

TOPAS has previously been shown to replicate the positron range and
distribution of positron energies at annihilation of TOF-PET
\cite{PET_2023_TMI_paper,TOPAS_Methods_paper} and serve as a reliable
simulation of other types of medical imaging procedures
\cite{TOPAS_proton_therapy,TOPAS_spectral_ct,TOPAS_dosimetry}.

\subsection{Gamma Ray Conversion Efficiency}
\label{conversion}
%
% Figure 5 Efficiency of 511 keV gamma conversion vs angle from normal in the phi direction (cylinder)
%
\begin{figure}[!th]
\centering
\includegraphics[angle=0,width=0.65\textwidth]{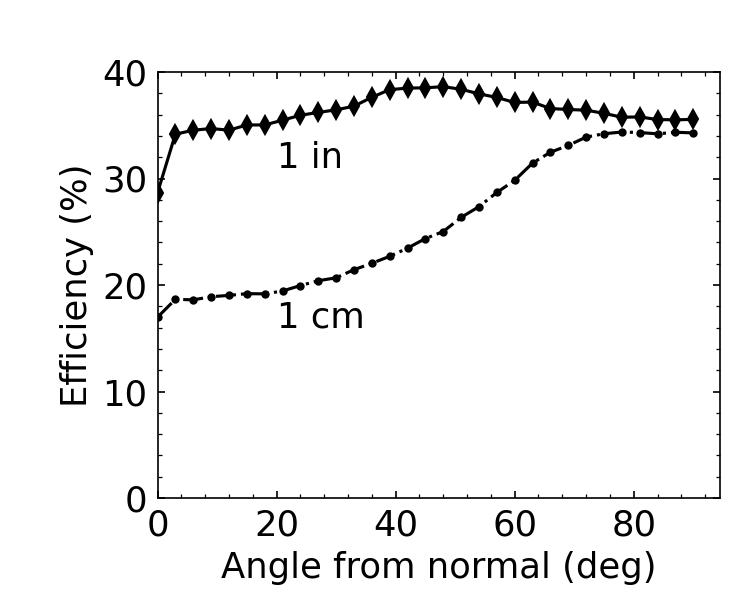}
\caption{The efficiency  for direct conversion of a 511 keV gamma ray
versus incident angle from the normal. Distributions are shown for LMCP
thicknesses of 1 cm and 2.54 cm (1 inch) with a lead-glass (``G4\_GLASS\_LEAD'') substrate.}
\label{fig:eff_by_ang_phi}
\end{figure}

Figure~\ref{fig:eff_by_ang_phi} displays the efficiency for MCP
thicknesses of 1 cm and 2.54 cm (1 inch) found in the TOPAS simulation
for direct conversion of a 511 keV gamma ray versus incident angle from
the normal to the lamina.  The efficiency includes the creation of a
primary electron that enters a pore by crossing a functionalized
pore-defining wall.

\subsection{Spatial Resolution}

%
% Figure 6 MGM_space_distribution -short direction across the pore
%
\begin{figure}[!th]
\centering
\includegraphics[angle=0,width=0.49\textwidth]{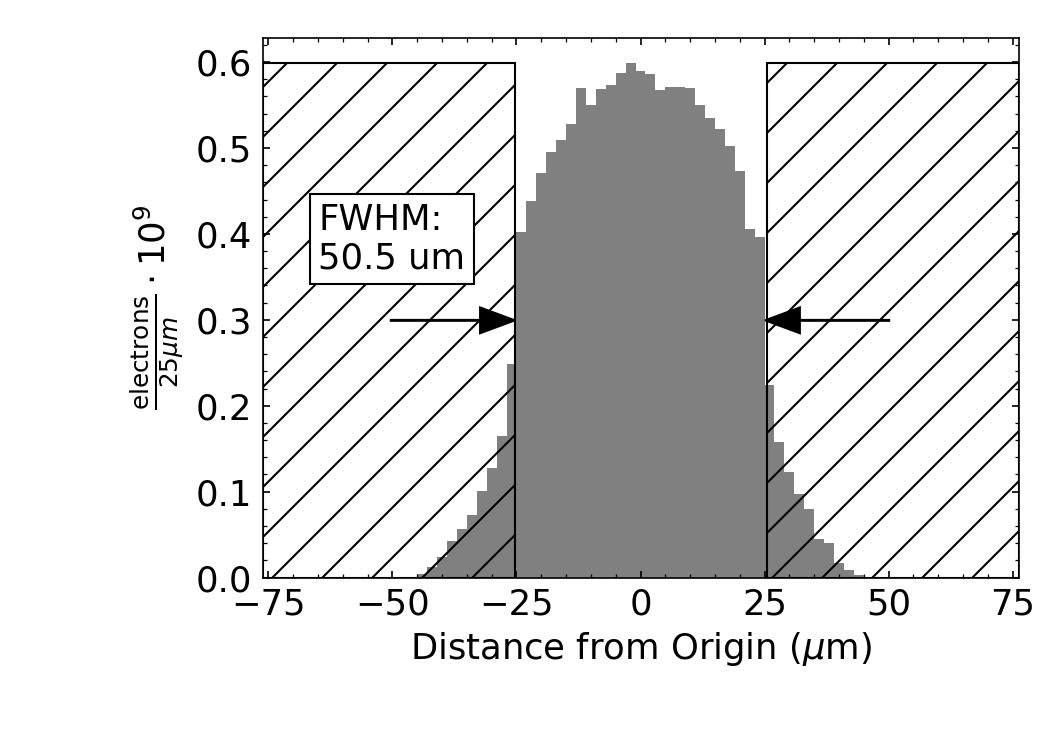}
\hfil
\includegraphics[angle=0,width=0.49\textwidth]{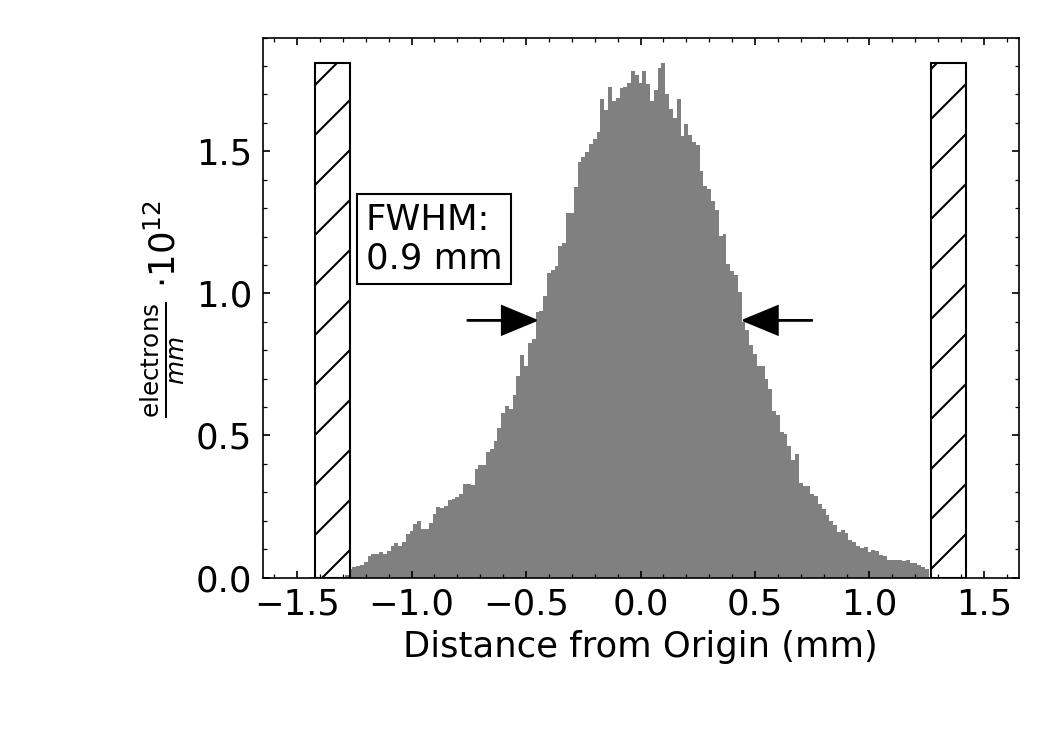}
\caption{The one-dimensional spatial distributions, in the `short' and
`long' dimensions, of charge at the exit of a 50 micron by 2.5 mm
rectangular pore as shown in Figure~\ref{fig:slab_w_ridges}. The pore
walls are indicated by hatching. The distributions are from a TOPAS
simulation of a single secondary electron created on the
`short'-dimension wall, 7.5 mm from the pore exit.}
\label{fig:MGM_channel_short_long_space_distribution}
\end{figure}

 Figure~\ref{fig:MGM_channel_short_long_space_distribution}
shows the simulated spatial distributions in the `short'  and `long'
dimensions of charge generated by a single secondary electron  created
on the `short'-dimension wall, 7.5 mm from the pore exit. The pore has
a uniform profile with transverse dimensions of 50 $\mu$m by 2.5
mm.\footnote{Here we have chosen to display an atypically large value
for one dimension to demonstrate the effect. The size of the pores in
an HGMT will most likely be in the tens of microns or less in both
dimensions.} The TOPAS simulation of secondary emission and
multiplication is initiated at the point of a primary electron on the
wall of the pore (left-hand surface in the Figure). The hatched regions
represent the substrate walls between neighboring pores.

\subsection{Time Resolution}

 The simulation of the time
resolution of an HGMT depends on the choice of many parameters of the
HGMT construction, including details of the materials and shapes of the
pore-forming surfaces, and consequently is beyond the scope of our
simulations. Data from a physical LMCP are essential in narrowing the
options towards high-resolution and robustness. In consequence, in the
simulations of imaging presented in
Section~\ref{whole_body_scanner_simulation_results}  we have used a
parametric approach, setting
the time resolution in the simulation to 100 ps independent of the
position of the gamma ray conversion in the pore. Appendix A presents
images simulated at different resolutions, including one using no TOF
information, and discusses possible future strategies for lowering the
spread in times due to the variation in conversion point to below 100
ps.

%
% Figure 8 HGMT wholebody
%
\begin{figure}[!th]
  \centering
  \includegraphics[angle=0,width=0.507\textwidth]{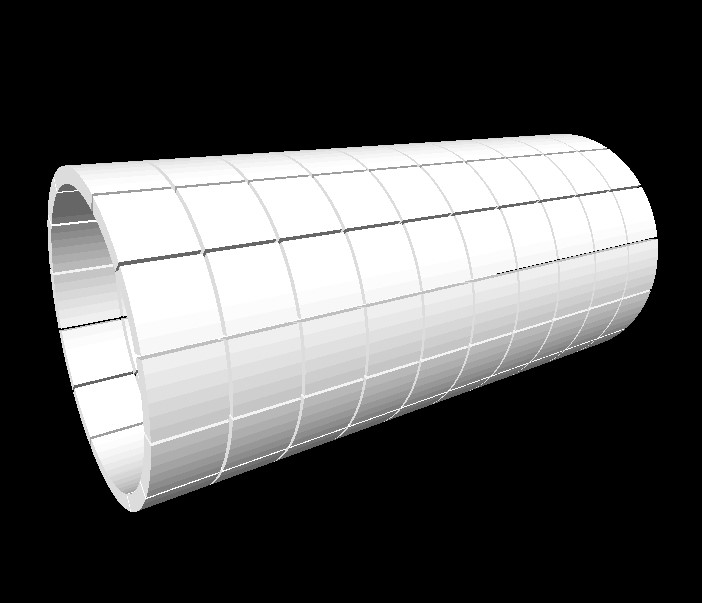}
  \hfil
  \includegraphics[angle=0,width=0.477\textwidth]{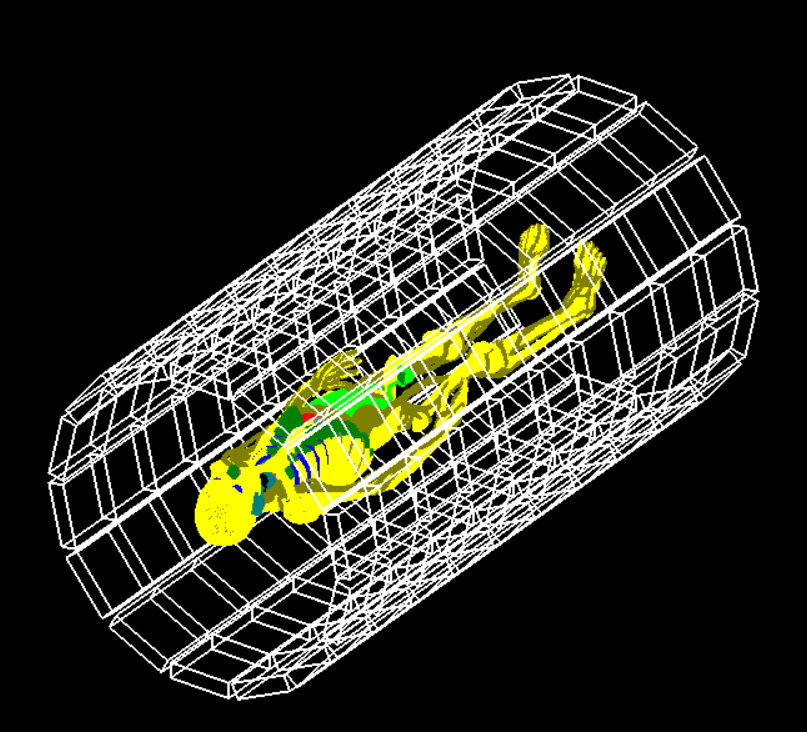}
  \caption{Left: A schematic whole-body PET scanner based on curved HGMT modules.
   Right: the graphical rendition of the whole-body scanner based on rectangular HGMTs including the XCAT phantom used in the
  TOF-PET simulation. The scanner is 200 cm long and has a bore radius of
  45 cm.} \label{fig:whole_body}
  \end{figure}

\section{Simulation of a Whole-Body HGMT TOF\--PET \\ Scanner}
\label{whole_body_scanner_simulation_results}

\subsection{Whole-Body Scanner Configuration}

The left-hand panel of Figure~\ref{fig:whole_body} shows a
representative whole-Body TOF-PET scanner made with curved~\cite{MGM_NIM_paper} HGMT modules. The
scanner benefits from the absence of a layer of scintillator to convert
the gamma rays to optical photons, and the absence of the cor\-re\-spond\-ing
photodetector system to convert the optical photons
to electrons. The right-hand panel shows the XCAT phantom inside the scanner as simulated in TOPAS. The scanner is 200 cm long and has a bore radius of 45 cm.

As the detector  typically would be $\approx5$ cm thick in the radial
direction, the HGMT facilitates integration into multi-modality systems
such as PET/MRI and PET/CT. The absence of the scintillator and
photodetector systems also substantially reduces complexity.

\subsection{Simulation of the Derenzo Phantom at Reduced Doses}

%
% Figure 9  Derenzo phantom at 1/100 dose
%
\begin{figure}[!th]
\centering
 \includegraphics[width=0.49\textwidth]{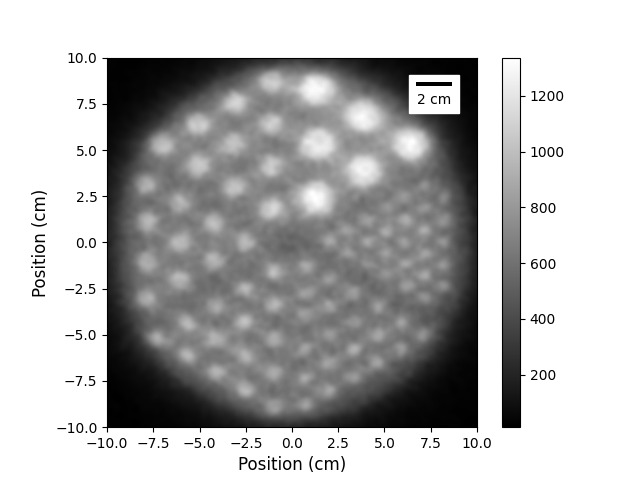}
\hfil
 \includegraphics[width=0.49\textwidth]{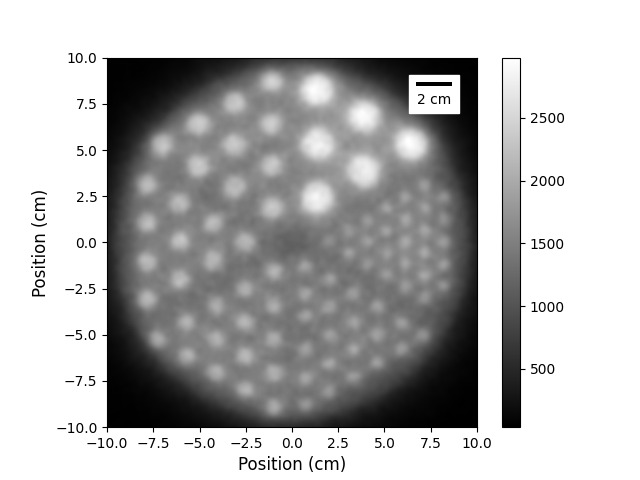}
 \caption{ Reconstructed images of the Derenzo phantom
at a dose of 150 Bq/mL for the rods and 50 Bq/mL for the background, a
factor of 100 lower than the benchmark dose. The thickness of the
converter LMCP is 1 cm in the left-hand image, and 2.54 cm (1 inch) in
the right-hand image.}
 \label{fig:Derenzo_lowdose_100}
\end{figure}

\begin{figure}[!th]
\centering
 \includegraphics[width=0.49\textwidth]{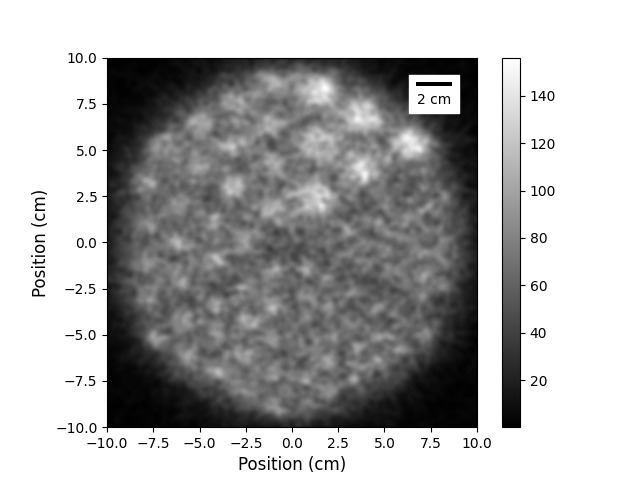}
\hfil
 \includegraphics[width=0.49\textwidth]{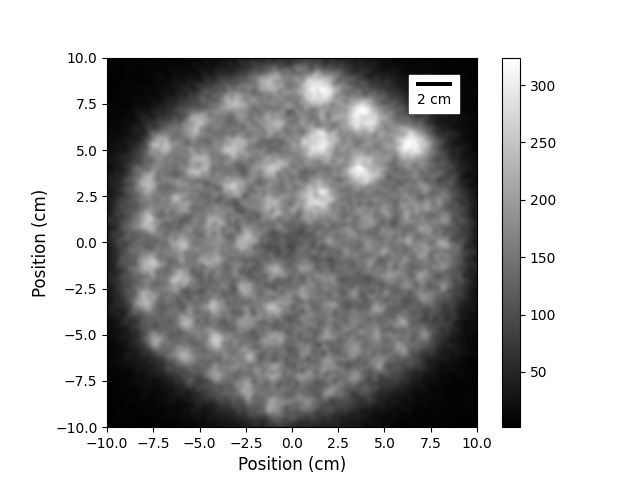}
 \caption{ Reconstructed images of the Derenzo phantom at a dose of 15
 Bq/mL for the rods and 5 Bq/mL for the background,
 a factor of 1000 lower than the benchmark dose.
 The thickness of the converter LMCP is 1 cm in the left-hand image, and
 2.54 cm (1 inch) in the right-hand image.   }
 \label{fig:Derenzo_lowdose_1000}
\end{figure}

Figures \ref{fig:Derenzo_lowdose_100} and \ref{fig:Derenzo_lowdose_1000} show reconstructed images of the Derenzo phantom \cite{Derenzo_phantom} at 1/100th and 1/1,000th reduced doses from a benchmark dose of 15 kBq/mL for the rods and 5 kBq/mL for the background at a scan time of 10 minutes. The timing resolution was taken as 100 ps
(FWHM); the spatial resolution in the plane of the LMCP at the pore
exit was conservatively set to 1 mm in both the `long' and `short'
dimensions.

%
% Figure 10  Derenzo phantom at 1/1000 dose
%

 \subsection{Simulation of the XCAT Brain Phantom and 2 cm-Diameter Lesion at Reduced Doses}
%
% Figure 11 XCAT brain at 1/100 dose
%
\begin{figure}[!th]
\centering
 \includegraphics[width=0.49\textwidth]{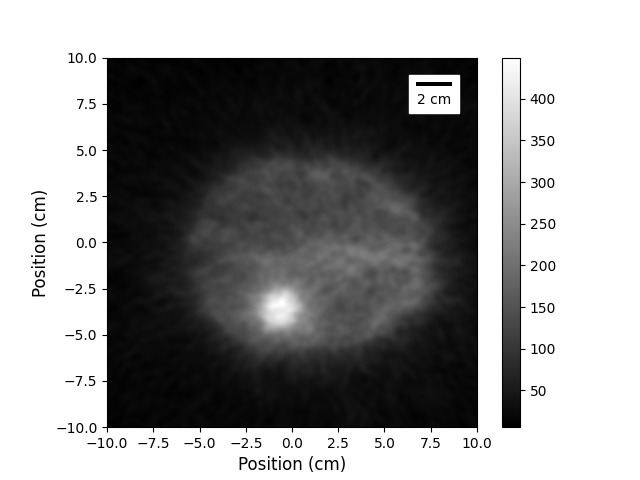}
 \hfil
 \includegraphics[width=0.49\textwidth]{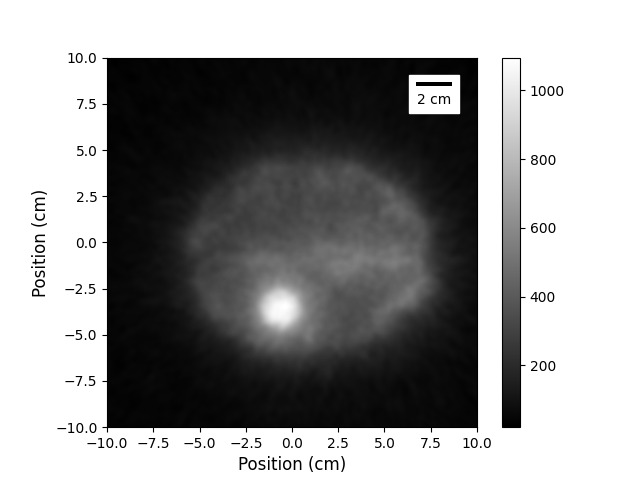}
 \caption{Reconstructed images of the XCAT brain phantom with a 2 cm-diameter spherical
lesion at a dose reduced by a factor of 100 from the
benchmark~\cite{XCAT_benchmark_dose}: 82.5 Bq/mL for white matter, 330
Bq/mL for gray matter, and 990 Bq/mL for the lesion. The thickness of the converter LMCP is 1 cm in the left-hand image, and
2.54 cm (1 inch) in the right-hand image. }
 \label{fig:XCAT_brain_lowdose_100}
\end{figure}

\begin{figure}[!t]
\centering
 \includegraphics[width=0.49\textwidth]{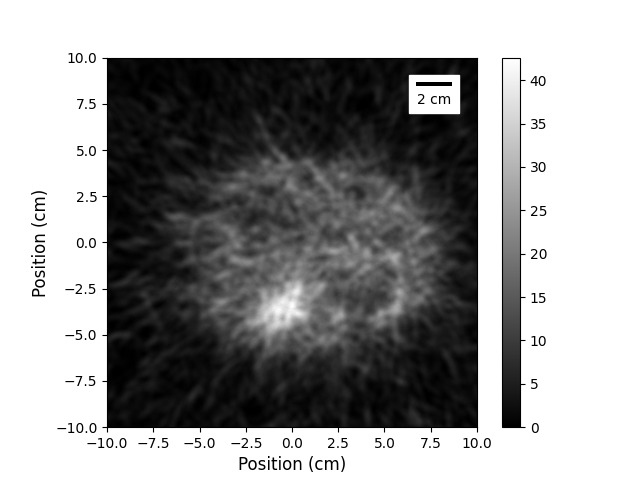}
 \hfil
 \includegraphics[width=0.49\textwidth]{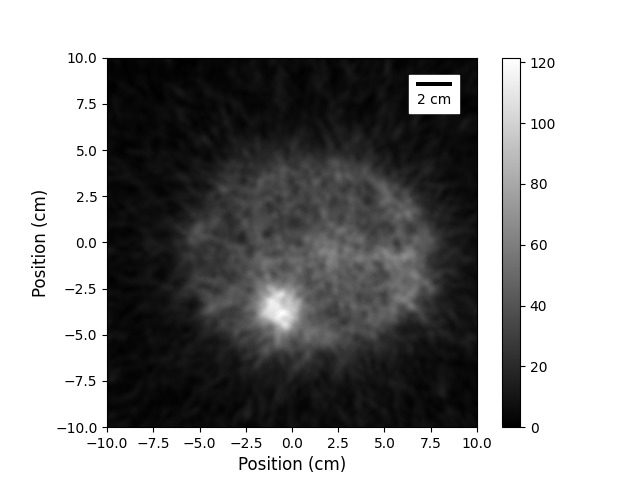}
 \caption{Reconstructed images of the XCAT brain phantom with a 2 cm-diameter spherical
lesion at a dose reduced by a factor of 1000 from the
benchmark~\cite{XCAT_benchmark_dose}:8.25 Bq/mL for white matter, 33
Bq/mL for gray matter, and 99 Bq/mL for the lesion. The thickness of the converter LMCP is 1 cm in the left-hand image, and
2.54 cm (1 inch) in the right-hand image. }
 \label{fig:XCAT_brain_lowdose_1000}
\end{figure}

\begin{figure}[!t]
\centering
 \includegraphics[width=0.49\textwidth]{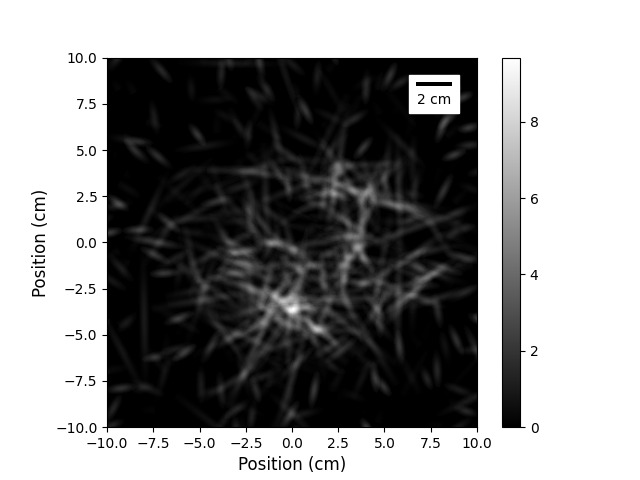}
 \hfil
 \includegraphics[width=0.49\textwidth]{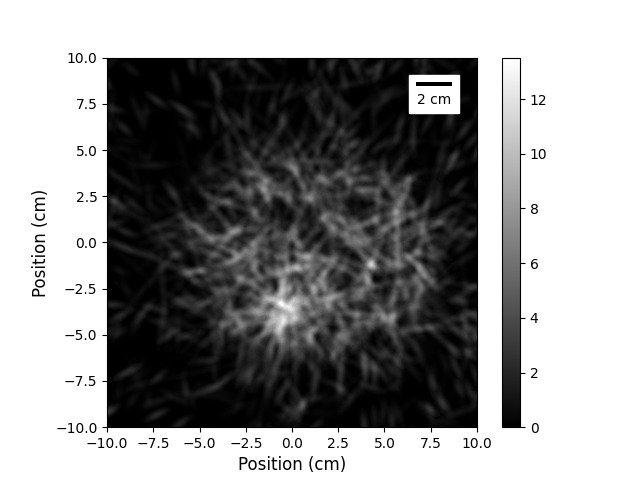}
 \caption{Reconstructed images of the XCAT brain phantom with a 2 cm-diameter spherical
lesion at a dose reduced by a factor of 10,000 from the
benchmark~\cite{XCAT_benchmark_dose}: 0.825 Bq/mL for white matter, 3.3
Bq/mL for gray matter, and 9.9 Bq/mL for the lesion. The thickness of the converter LMCP is 1 cm in the left-hand image, and
2.54 cm (1 inch) in the right-hand image. }
 \label{fig:XCAT_brain_lowdose_10000}
\end{figure}

The XCAT brain phantom~\cite{XCAT2010paper} was also simulated at 1/100th, 1/1,000th, and 1/10,000th reduced doses from a benchmark dose of 8.25 kBq/mL for white
matter, 33 kBq/mL for gray matter, and 99 kBq/mL for the spherical
lesion~\cite{XCAT_benchmark_dose} for an estimated 10-minute scan. Figures \ref{fig:XCAT_brain_lowdose_100}, \ref{fig:XCAT_brain_lowdose_1000}, and \ref{fig:XCAT_brain_lowdose_10000} show the brain at these doses, respectively. While the images at 1/10,000th dose
are not good enough for detailed diagnosis, they may be enough to
suggest a follow-up scan at a higher dose, and may inform strategies
for regular screening of appropriate populations, such as selective
annual exams for breast cancer.

%
% Figure 12 XCAT brain at 1/1000 dose
%

%
% Figure 1 XCAT brain at 1/10000 dose
%

\section{Portable and Animal TOF-PET Scanners}
\label{sec:portable-animal-pet}
%
%Figure 17 HGMT Portable PET
%
\begin{figure}[!th]
\centering
\includegraphics[angle=0,width=0.55\textwidth]{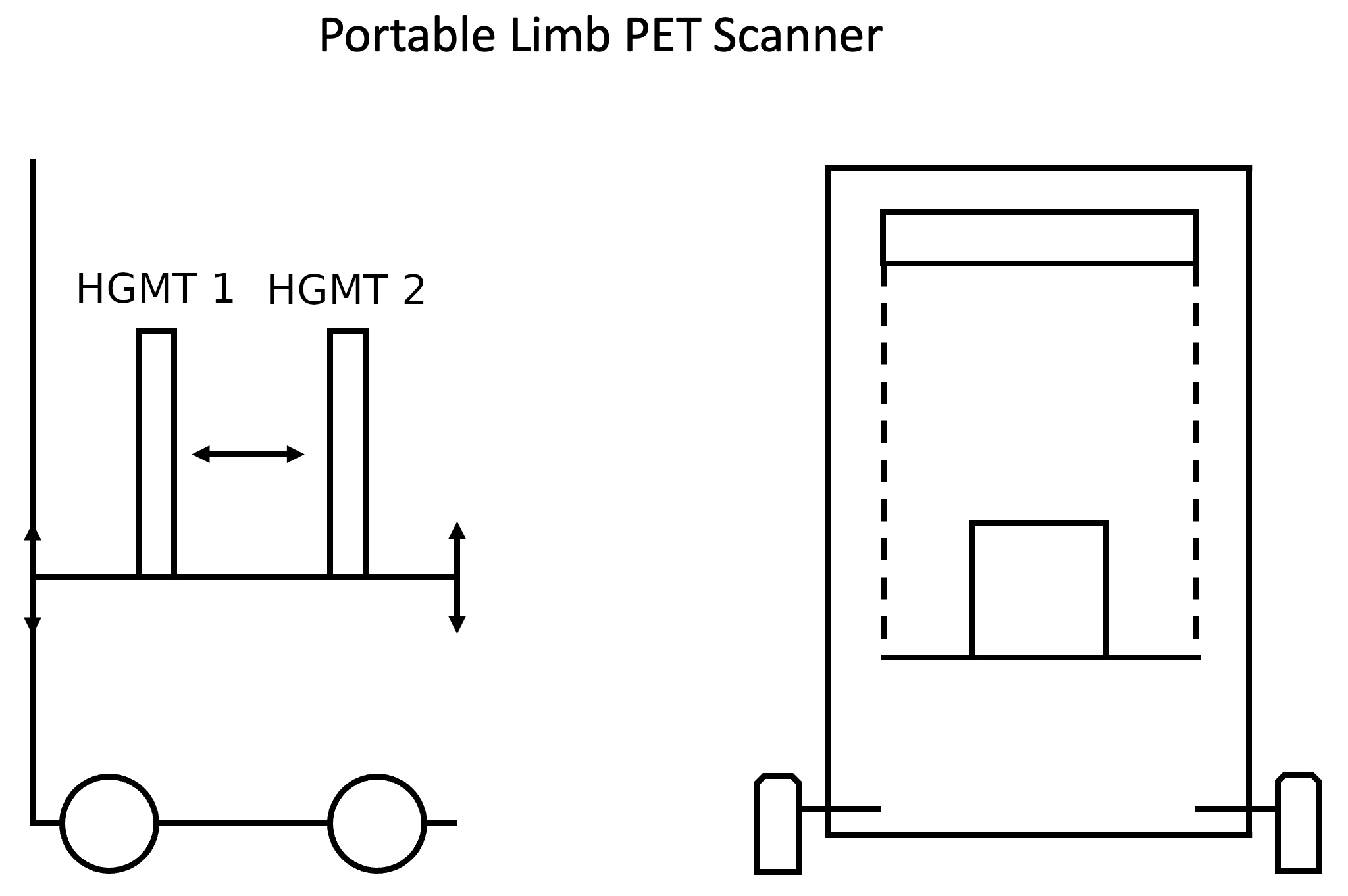}
\caption{A schematic portable PET scanner based on HGMT modules.
The scanner comprises several HGMTs. The HGMTs can be sealed or pumped by a small pump
on the cart. The HGMTs can sit on a stage adjustable in height and aperture for imaging
 breaks in arms or legs, for example. Local electronics and computation can give a real-time image.}
\label{fig:portable_PET}
\end{figure}

A significantly lower dose may allow the use of portable TOF-PET
scanners for clinical applications and facilities for which PET is not currently
possible or economical. Examples are hair-line fractures, for which the
current standard of X-rays has a significant rate of
non-detection~\cite{zero_for_four}, but for which PET has high
sensitivity. Figure~\ref{fig:portable_PET} shows a sketch of a simple
two-module portable scanner on a portable cart that can be adjusted in
height and aperture to accommodate legs and arms, for example. Because
the HGMT does not have a photocathode, it does not need ultra-high
vacuum (UHV), and can be pumped with a small vacuum pump located on the
cart. The system can be valved off, transported, and restarted at a new
location.

%Figure 18 HGMT animal
\begin{figure}[!th]
\centering
\includegraphics[angle=0,width=0.65\textwidth]{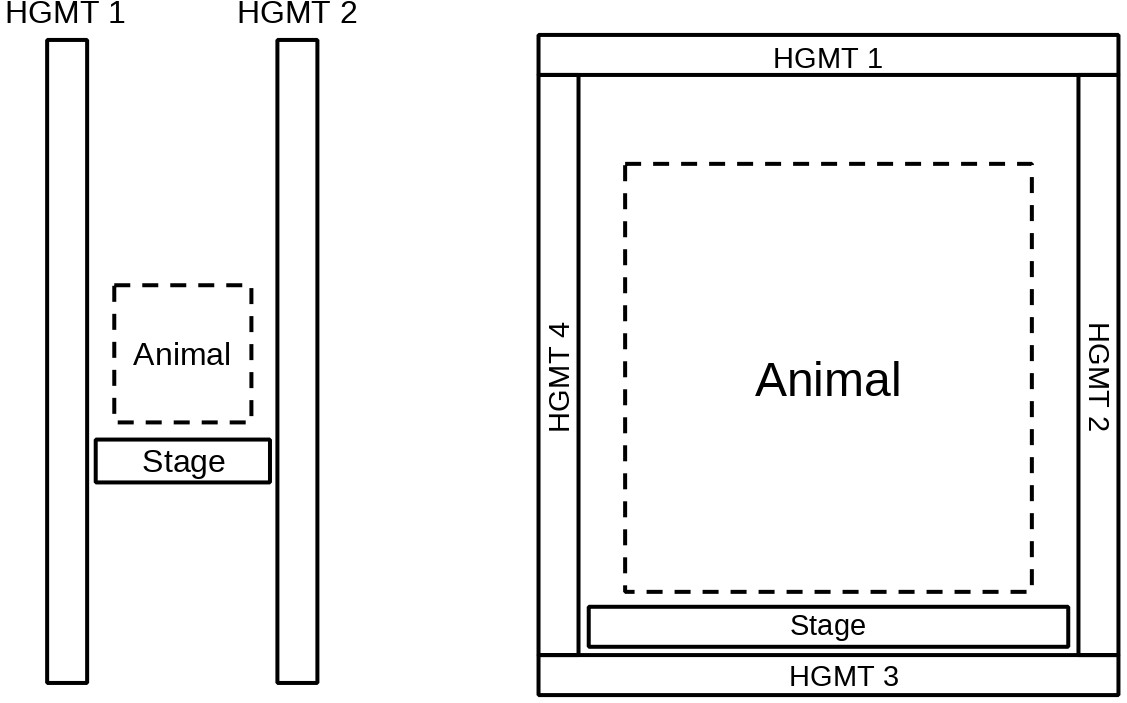}
\caption{Left: A schematic economical small-animal PET scanner formed
by two HGMT modules. Right: A small-animal PET scanner formed by four
modules in azimuth.}
\label{fig:small_animal}
\end{figure}

Figure~\ref{fig:small_animal} shows a representative PET scanner for
small animals. The large area of the HGMT may allow coverage of much of
the solid angle with only two HGMTs. An array of multiple HGMTs covering four or six sides can provide coverage for larger animals.

\section{Summary and Conclusions}
\label{summary_and_conclusions}

 We have adapted the TOPAS Geant4-based tool kit to simulate  surface
direct conversion in a Laminated Micro-Channel Plate (LMCP) constructed
from thin lead-glass laminae 150 microns-thick \cite{MGM_NIM_paper}.  An
LMCP 2.54 cm-deep  is predicted to have a $\ge 30$\% conversion
efficiency to a primary electron that penetrates an interior wall of a
pore. We present  space and time resolutions from the subsequent
secondary electron shower. Images from initial simulations of
whole-body HGMT TOF-PET scanners at doses reduced from literature
benchmarks by factors of 100 and 1000 are presented.

In whole-body PET scanners the technique eliminates the scintillator
and pho\-to\-de\-tec\-tor subsystems. In addition, the absence of a
photocathode eliminates many onerous aspects of current UHV
fabrication, as it allows assembly of large arrays at atmospheric
pressure with less stringent vacuum requirements, including use of
pumped and cycled systems.

TOPAS simulations of the Derenzo and XCAT-brain phantoms are
presented with dose reductions of factors of 100 and 1000 from
literature benchmarks. Benefits of such reduction would
include routine screening for early tumor detection, use of
PET for pediatric diagnostics, and a larger installed facility
base in rural and under-served populations.

Application-specific implementations of the surface direct production
technique em\-ployed in the HGMT are also candidates for large-area
arrays for use in detectors across a wide range of fields in physics.

In conclusion, initial TOPAS Geant4-based simulation studies of
whole-body TOF-PET using direct conversion of the gamma rays to
electrons via the Photoelectric and Compton Effects indicate the
possibility of useful imaging at substantially lower radiation doses.
The LMCP technique of laminated construction of micro-channel plates,
in this case with at least part of each lamina consisting of a material
with heavy nuclei such as those of lead or tungsten, would allow access
to many operational parameters for detector optimization. We hope
others interested in making the unique capabilities of
positron emission tomography  widely and routinely available will join
us in applying resources to building and testing hardware.

%

% ================================ APPENDIX A ============================
%

\section{Appendix A: Time Uncertainty From the Varying Conversion Point}
\label{appendix_A}

\begin{figure}[!th]
  \centering
   \includegraphics[width=0.49\textwidth]{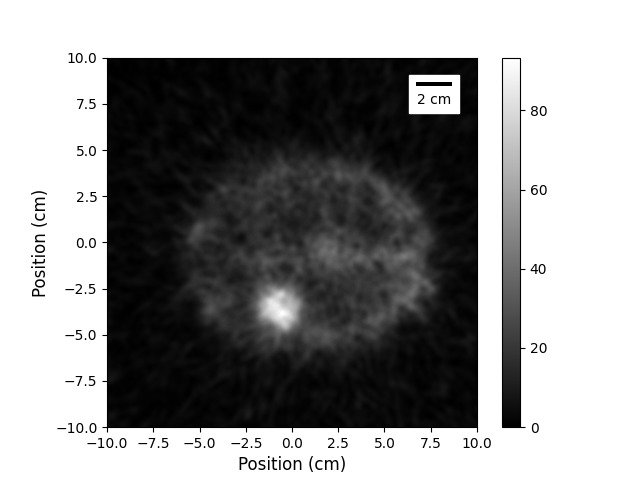}
   \hfil
   \includegraphics[width=0.49\textwidth]{brain_c_1in_100ps_1mm_1_000th_v2a.jpg}
   \hfil
   \includegraphics[width=0.49\textwidth]{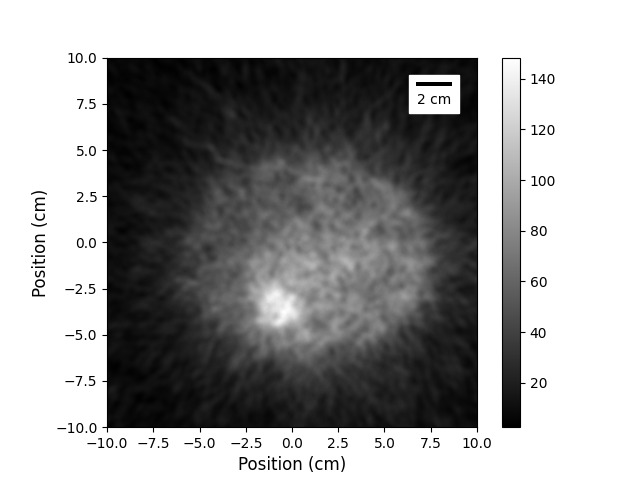}
   \hfil
   \includegraphics[width=0.49\textwidth]{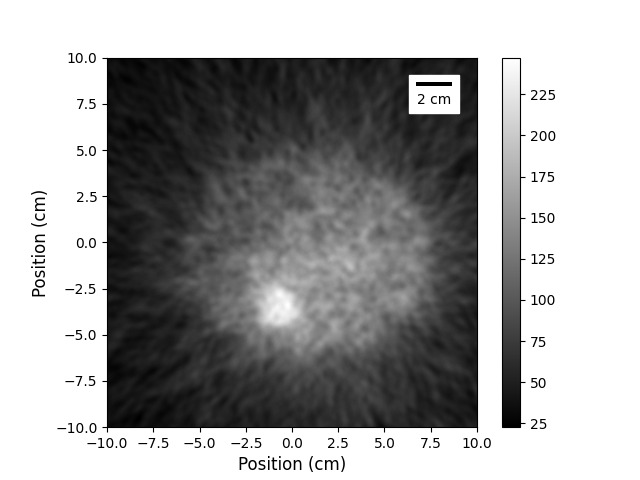}
   %brain_1inHGM_200psFWHM_1mmSigma_1_000th_v1a.jpg}
   \caption{Reconstructed images at different time resolutions
   of the XCAT brain phantom with a 2 cm-diameter spherical
  lesion at a dose reduced by a factor of 1000 from the
  benchmark~\cite{XCAT_benchmark_dose}: 8.25 Bq/mL for white matter, 33
  Bq/mL for gray matter, and 99 Bq/mL for the lesion. Top left: Time
  resolution of 50 ps (FWHM); Top right: Time resolution of 100 ps
  (FWHM); Bottom left: Time resolution of 200 ps (FWHM); Bottom right:
  Time resolution of 10,000 ps (FWHM), equivalent to no time cut. }
   \label{fig:XCAT_brain_lowdose_1000_4time_resolutions}
  \end{figure}

%\subsection{The Goal of Time Resolutions Comparable to Transverse Spatial Resolutions}
 The power of measuring the position of the $e^+e^-$
annihilation as a point in three dimensions, rather than the
2-dimensional Line-of-Response (LOR), represents a reduction of
orders-of-magnitude in required radiation dose, as well as a move from
image reconstruction to pattern recognition. There are now efforts to
achieve a resolution in the longitudinal direction of the LOR
comparable to the transverse resolution\footnote{We note that 10 ps
corresponds to a light travel distance of 3 mm for a single-ended
measurement in vacuum.} of a few mm or
less~\cite{LeCoq_2019_case,LeCoq_2020_10ps_challenge,PET_2023_TMI_paper}.
The studies of signal development and wave-form sampling front-end
electronics for the requisite time resolution are quite
mature~\cite{Ritt_anode_Genat_anode_drake_delagnes_2011_workshop,Ritt_limitations}
and so the problem devolves to the time-resolution of the detector
itself. The dependence of the image on the time-of-flight resolution is
illustrated in
Figure~\ref{fig:XCAT_brain_lowdose_1000_4time_resolutions}, which shows
the simulated image of the XCAT brain at a dose reduced by a factor of
1000 for four different time resolutions (FWHM): 50, 100, 200 ps, and
10 nsec, i.e. a resolution much larger than the scanner.
%
%  Still Appendix A
%  Figure 15 XCAT brain at 1/1000 dose and 50, 100, 200, and 10,000 ps time resolutions
%

\subsection{Contribution to the Timing Resolution from the Secondary
Shower Formation}

\begin{figure}[!th]
\centering
\includegraphics[angle=0,width=0.49\textwidth]{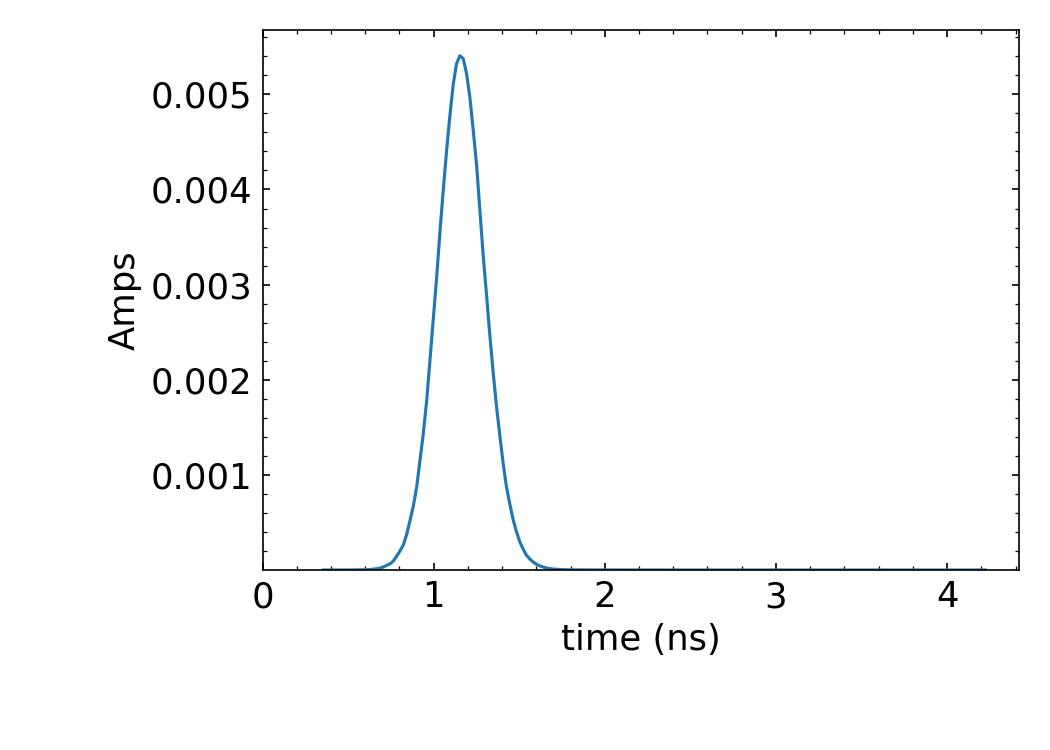}
\hfil
\includegraphics[angle=0,width=0.49\textwidth]{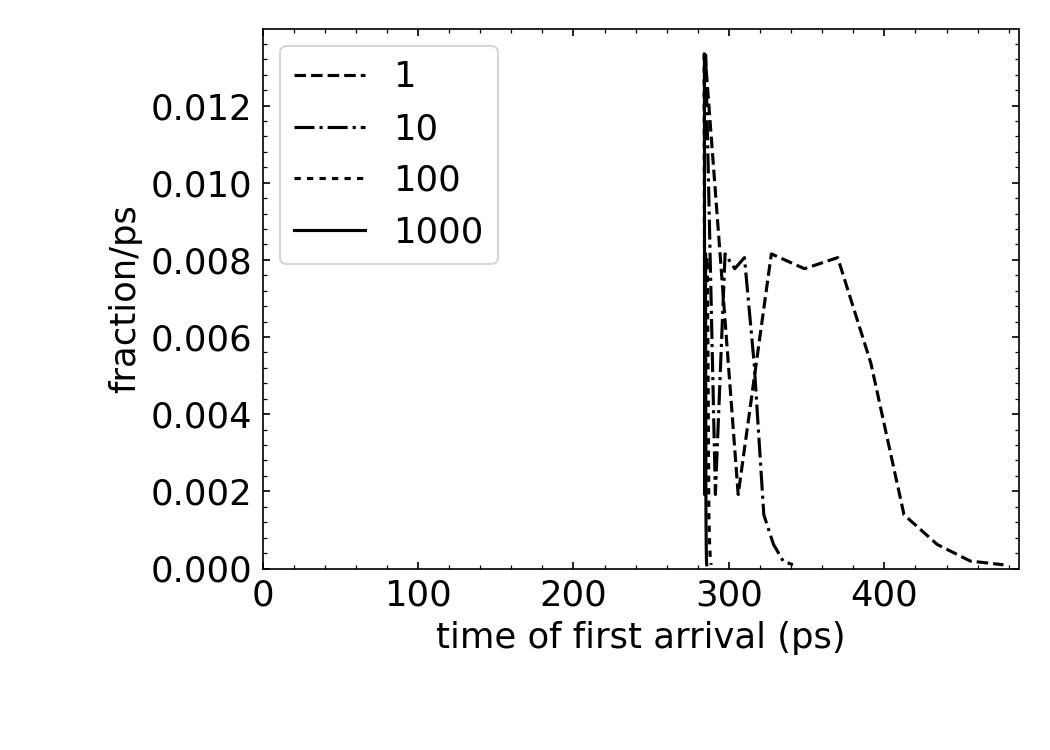}
\caption{Left: The current  predicted by the TOPAS simulation in the
secondary electron shower at the pore exit versus time for a cascade
initiated by a single secondary electron 7.5 mm from the exit end of
the pore and centered on one of the short sides. Right: The  time of
the first secondary electron that arrives at the exit plane of the
lead-glass converter LMCP from a TOPAS simulation of the electron
shower in one pore of the LMCP of Figure~\ref{fig:slab_w_ridges},
versus the number of secondary electrons created by the primary
electron from the direct gamma conversion.}
\label{fig:current_first_one_in}
\end{figure}

The contribution to the timing resolution from the secondary shower
development will be dominated by the multiplicity early in the shower,
in particular the multiplicity of the secondary electrons created by
the primary electron crossing the functionalized pore-forming surfaces.
 The left hand panel in Figure~\ref{fig:current_first_one_in}
 shows a simulated pulse shape
from  a single secondary electron 7.5 mm from the exit end of the pore
and centered on one of the short sides. The right-hand panel presents
simulations of showers started from a different multiplicity of
secondary electrons. Poisson statistics predicts that a higher
multiplicity of initial secondary electrons improves the time
resolution as the probability of seeing no electrons in the initial
time interval falls exponentially in the expected number per unit time
with the length of the interval.\footnote{We call the case in which the
system has enough gain that a single electron triggers the system to
determining the time  `First-One-In'.} In the simulation of
Figure~\ref{fig:current_first_one_in}, electrons are started
one-at-a-time 7.5 mm from the exit of the pore, centered in the short
direction between the two walls. Any electron exiting the plane enters
the amplifying LMCP(s) below, followed by an anode.

%
% Figure 16 Distributions of current and first-one-in-times for MGM (Appendix A)
%

The contribution to the time resolution from the secondary shower in
the limit of large first-strike secondary emission, shown in
Figure~\ref{fig:current_first_one_in}, is well below 50 ps. The
resolution will consequently be dominated by the varying distance of
the start of the shower to the pore exit.

\subsection{Contribution to the Timing Resolution from the Starting \\ Position Along the Pore}

\begin{figure}[!h]
\centering
\includegraphics[angle=0,width=0.80\textwidth]{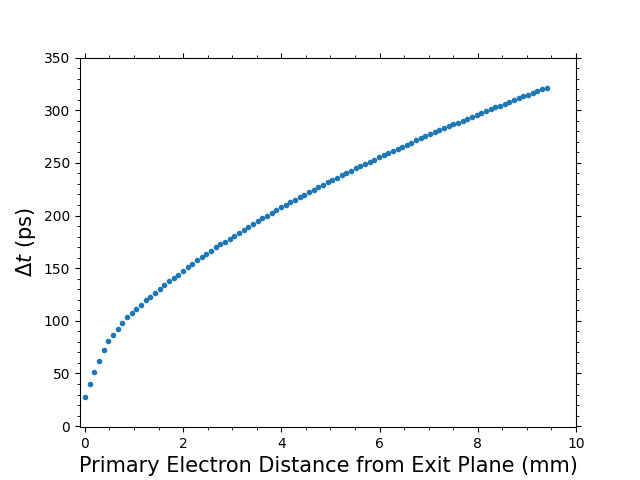}
\caption{The  time of arrival of the first electron at the exit plane
versus the height of the initiating point in the LMCP pore wall.}
\label{fig:time_offset_vs_primary_position}
\end{figure}

Figure~\ref{fig:time_offset_vs_primary_position} shows the time of
arrival of the first secondary electron at the exit plane of the
lead-glass converter LMCP versus the height of the initiating point in
the pore wall. For a given desired time resolution, the slope
determines the maximum thickness in the LMCP of the lead-glass
converter section before the conventional MCP amplification section of
the HGMT for this contribution not to dominate in the absence of
other solutions.

%
% Figure 17 Time of arrival versus the height in the pore wall (Appendix A)
%

%\clearpage

\subsection{Can One Overcome the Position Dependence in the LMCP Pore?}

%
%
% Figure 18 Multiple HGMT Sub-units (Appendix A)
%
\begin{figure}[!ht]
\centering
\includegraphics[angle=0,width=0.75\textwidth]{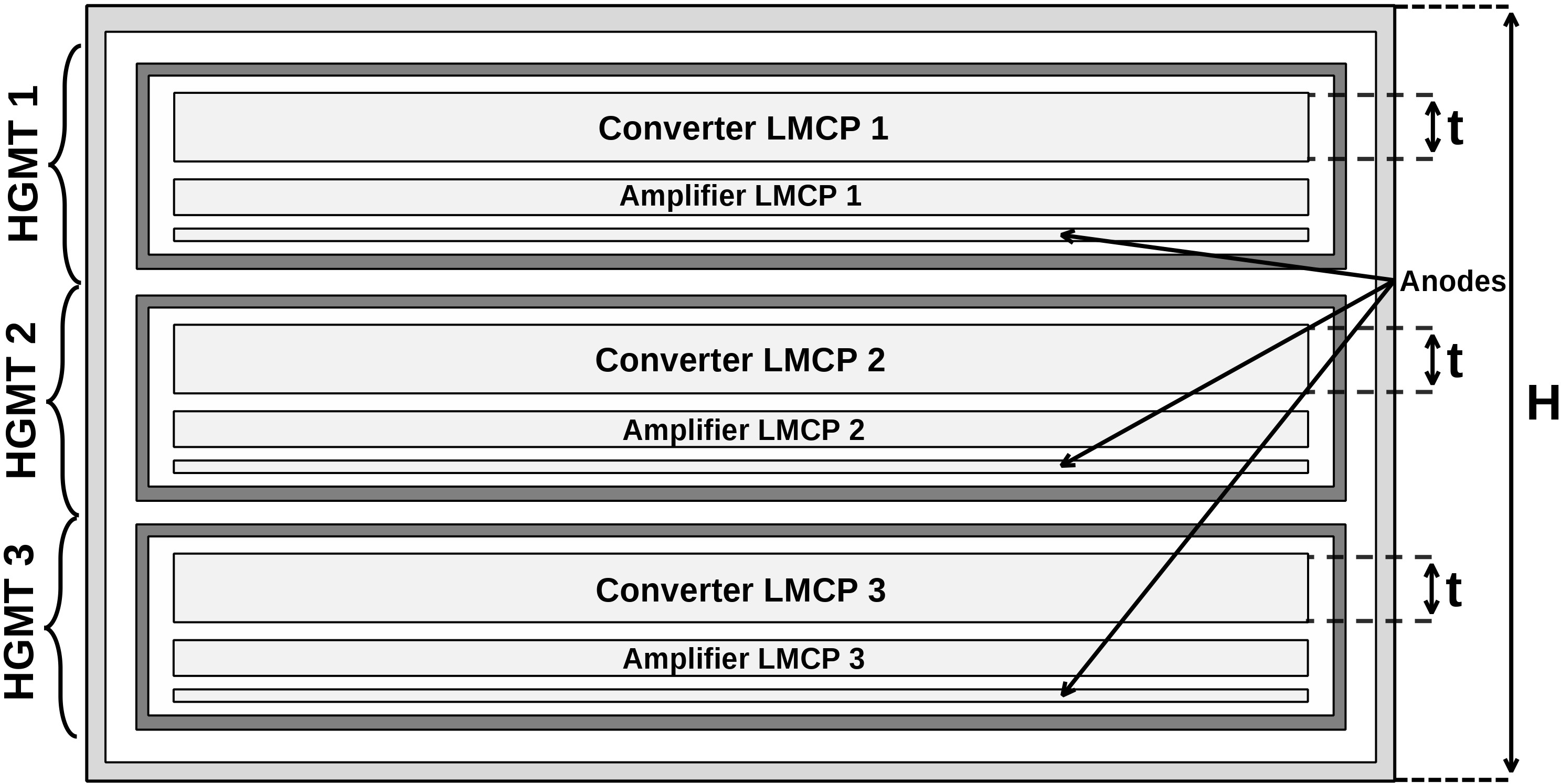}
\caption{An example of using stacked HGMT internals with thinner
conversion LMCPs to lower the uncertainty due to the primary
interaction in the pore walls.} \label{fig:HGMT_stack}
\end{figure}

 One straight-forward strategy to lower the time resolution
below 100 ps is to stack multiple HGMT internal modules, each
consisting of a converter LMCP made with lead-glass followed by one or
more amplification sections made from B33 glass or equivalent, and an
anode. A stack of these HGMT sub-modules with a total conversion path
length of a several cm will be less expensive  and less bulky than
conventional crystal systems.\footnote{Strip-line anodes are
constructed from inexpensive two-layer printed circuit boards, and 150
electronics channels can cover a square meter. The sub-modules share a
common hermetic package and High Voltage  distribution.}
Figure~\ref{fig:HGMT_stack} shows an example stack of 3 HGMT assemblies
in a common vacuum package. Each of the Conversion LMCP/Amplification
LMCP/anode assemblies would provide a conversion path length
appropriate for the desired time resolution and overall conversion
efficiency.  The system cost and bulk are expected to be less than
those of conventional crystal-based scintillator systems.

The LMCP technique may support more-sophisticated strategies than the
above brute force stacking, and can be explored in a program of
measurements of actual devices. For example, the design could enhance
the correlation of pulse height with conversion point in the LMCP. A
pore design with discrete strike points spaced at small intervals, for
example 1 mm, and resistances such that the potential between them is
well up on the SEY curve (e.g. at 4 for MgO)~\cite{Slade_SEY_NIM} will
produce some measure of discrete gain versus height from the pore exit.
A more speculative example to be explored that illustrates the
flexibility of the open-face LMCP construction technique is to
incorporate RF pickups at intervals along pores before assembly. The
pattern of signals from the antennae along the pore would locate the
start of the secondary shower. However the brute force approach of
multiple thin LMCP assemblies described above should be feasible and
affordable.

%\clearpage
\section*{Acknowledgments}
We thank Joseph Perl and Paul Segars for the exemplary development of
TOPAS and XCAT and for their remarkable user support. We are indebted
to Mary Heintz for exceptional computational system development and
advice. Benjamin Cox provided highly informed encouragement, advice,
and support. We are grateful to Ian Goldberg, Justin Gurvitch, and
Richmond Yeung for graphics contributions.  We deeply thank an
anonymous reviewer for thoughtful comments.

K. Domurat-Sousa, and C. Poe were supported by the University of
Chicago College, Physical Sciences Division, and Enrico Fermi
Institute, for which we thank Steven Balla and Nichole Fazio, Michael
Grosse, and Scott Wakely, respectively. C. Poe was additionally
supported by the University of Chicago Quad Undergraduate Research Scholars program and the Jeff Metcalf Internship program.

\enlargethispage{6.0in}
%
%============================= END APPENDIX A =====================================
%==================================================================================

\newpage
\clearpage

%===============================BEGIN BIBLIOGRAPHY===========================================================

% THE BIBLIOGRAPHY


\begin{thebibliography}{99}
\bibitem{zero_for_four} One of the authors (HJF) observed non-detection
by X-ray for four 90-year-old women who each spent a week in the
hospital before a PET scan produced an unmmistakable diagnosis. You can
calculate the efficiency from Poisson statistics yourself.
\end{thebibliography}
  \end{document}